\documentclass[12pt]{JHEP3}

\usepackage{graphicx}
\usepackage{amsfonts, amsmath}
\usepackage{mathrsfs}
\usepackage{setspace}
\usepackage[utf8]{inputenc}
\usepackage{comment}
\usepackage{cite}

\newcommand{\be}{\begin{equation}} 
\newcommand{\ee}{\end{equation}}

\newcommand{\lie}[1]{{\mathscr L}_{\raisebox{-1mm}{$\scriptstyle #1$}}}

\title{Kerr-CFT From Black-Hole Thermodynamics}

\author{Bruno Carneiro da Cunha$^{a}$,
  Amilcar R. de Queiroz$^{b,c}$\footnote{bcunha@df.ufpe.br,
amilcarq@unb.br},
\\
$^a$Departamento de Física, Universidade Federal de Pernambuco,
53901-970, Recife, Pernambuco, Brazil
\\
$^b$Instituto de Física, Universidade de Brasília, Caixa Postal
04455,
70919-970, Brasília, DF, Brazil
\\
$^{c}$International Center for Condensed Matter Physics (ICCMP),
Universidade de Brasília, Caixa Postal 04667, Brasília, DF,
Brazil\\
}

\abstract{We analyze the near-horizon limit of a
general black hole with two commuting killing vector fields in
the limit of zero temperature. We use black hole thermodynamics methods
to relate asymptotic charges of the complete spacetime to those
obtained in the near-horizon limit. We then show that some
diffeomorphisms do alter asymptotic charges of the full spacetime, even
though they are defined in the near horizon limit and, therefore, 
count black hole states. We show that these conditions are
essentially the same as considered in the Kerr/CFT corresponcence. From
the algebra constructed from these diffeomorphisms, one can extract its
central charge and then obtain the black hole entropy by use of Cardy's
formula.}

\keywords{Kerr-CFT, Extremal Black Hole, Duality, Virasoro Algebra, Central Charges}

\preprint{\today}

\begin{document}


\section{Introduction}

Central charges of symmetry algebras are helpful to either organize the
dynamics of a classical system or the spectrum of a quantum one
\cite{Arnold}.
In three dimensional gravity, a  particularly relevant set of charges
found by Brown and Henneaux \cite{Brown1986b} allows
for an interpretation of the states of the theory as
diffeomorphisms of space-time. In that particular example, the states
arise as coordinate transformations which leave the asymptotic
structure of ${\rm AdS_3}$ invariant but do modify the subleading
terms and thus change the conserved quantities. Assuming that the
underlying theory is unitary, one can then use generic results in
two-dimensional conformal field theory \cite{Cardy1986} to count the
number of states in the spectrum. This in turn paved way
for the Strominger-Vafa calculation in string theory
\cite{Strominger:1996sh}, where the same 
Virasoro algebra was found from the string excitations of the $D1-D5$
system. Then the appearance of classical central charges became an
indication that one could understand the entropy of black holes, or
gravitational systems in general, as somehow counting the number of
diffeomorphically related but inequivalent metrics.

In Guica et {\it al.} \cite{Guica:2008mu}, an interesting example of
the procedure outlined above was put forward for the extremal Kerr
black hole. The really interesting feature of the correspondence in
that case is that it does not apparently rely on supersymmetry.
Instead, the idea is based on the fact that, in the case of the
extremal Kerr black hole, the near-horizon limit has as
isometry the
algebra $SL(2,\mathbb{R})\times U(1)$, as remarked by Bardeen in the
70's (for a more recent application, \cite{Bardeen:1999px}).
An appropriate choice of
boundary conditions of this near-horizon geometry
$AdS_2\times S^2$ allow them to show that the $U(1)$ sector of
the isometry
algebra can be enhanced to a Virasoro algebra with central charge
$c=12 J/\hbar$. From this algebra Cardy's formula was used to
compute the entropy, assigning to the extremal horizon the Frolov
temperature \cite{Frolov1989}. The result, $S=\pi^2 c
T/3=2\pi J/\hbar$, exactly reproduces the
celebrated Bekenstein-Hawking entropy of the extremal Kerr black hole.

There were many works applying the procedure to other
examples of extremal spinning black holes. A partial list include:
Kerr-Bolt Spacetimes \cite{Ghezelbash2009c}, extremal Kerr-Sen black
hole that appears as solutions in the low energy limit of heterotic
string theory in 5D \cite{Ghezelbash2009a}, extremal rotating
Kaluza-Klein black holes \cite{Azeyanagi:2008kb}, D1-D5-P and the BMPV
black holes \cite{Azeyanagi:2008dk,Isono2008}, general rotating black
hole solutions in gauged and ungauged supergravities
\cite{Chow2008,LuJHEP0904:0542009}, and solutions of higher order
theories of gravity \cite{Krishnan:2009tj}. Furthermore, by considering
different boundary conditions another realization of the Virasoro
algebra of the above procedure was also considered in
\cite{Matsuo2009,Matsuo2009a}. A partial classification of distinct
boundary conditions are discussed in \cite{Rasmussen2009b}.

As stated above, a key step in \cite{Guica:2008mu} is the
appropriate choice of the boundary conditions. In their work, they use
Barnich and Brandt \cite{Barnich:2001jy} formulas for the central
charges, and by {\it a posteriori} inspection of these formulas they
were able to discover boundary conditions allowing for Virasoro algebra.
Now, it is certainly desirable to be able to obtain these boundary
conditions via more physical arguments.

In the present work, we give some arguments on some necessary physical
conditions that leads to these boundary conditions. For that we pose
the near-horizon limit in an interesting form (\ref{eq:scale-transf}). That
allows
us to see this near-horizon limit as a decoupling limit, and therefore
understand it as a renormalization group flow, typical of
gauge/gravity dualities, in a still-to-be-found dual gauge theory. We
next apply this renormalization group flow reasoning to relate
observables in the Kerr space-time to (usual infinite) near-horizon
observable. In particular, we apply this to the case of the mass,
obtaining thus condition \ref{eq:scaling-prescription}. From this condition, it
is
straightforward to obtain the boundary conditions \cite{Guica:2008mu}.

The present work is organized as follows: in Section 2, we present
some general geometric conditions for a space-time to yield a (fibered)
$AdS_2\times X$ geometry in the near-horizon limit; in Section 3, we
use the zeroth law of black hole thermodynamics and the extremality
condition to obtain the near-horizon limit. We then write and discuss
the near-horizon limit as a decoupling limit; in Section 4, boundary
conditions for the near-horizon geometry that leads to a Virasoro
algebra are obtained by using the first law of black hole
thermodynamics. We then argument that this boundary conditions can be
obtained by relating full space-time observables to near-horizon
observables; in Section 5, we obtain the central charge of these
Virasoro algebra and use it to obtain the microscopic entropy of the
extremal spinning black holes. We close this letter with concluding
remarks.

\section{Two commuting Killing fields}

There are many examples of space-times that allow for a fibered ${\rm
  AdS_2}\times X$ description in the near horizon limit. In fact, the
program began almost forty years ago \cite{Bardeen1971}. In all
examples, one starts from a regular solution of gravity with a number
of symmetries and by a scaling limit, enhances the symmetry to ${\rm
  SL(2,\mathbb{R}})$, that of ${\rm AdS_2}$. The scaling limit is
taken by directly choosing a set of coordinates, and although some of
the coordinates have a direct geometric interpretation (like $t$),
some do not (like $r$). In the spirit of general covariance, we will
set geometrically the conditions under which the near-horizon limit of
a black hole will yield a fibered ${\rm AdS_2}\times X$ geometry.

We start with a solution allowing two commuting Killing vector fields. A time
translation $\partial/\partial \bar{t}$ will give us a definite global energy
and a rotation $\partial/\partial \bar{\phi}$ will yield one combination
of the angular momenta entering the first law.
Both generators commute between themselves. Indeed, the action of either
generator will keep invariant the conserved charge associated with the other
generator. Following the discussion in Sec. 7.1 of \cite{Wald}, we choose
coordinates $\bar{t},\bar{\phi},x^i$, such that the metric makes full use of the
symmetries.

Obviously, the metric components are functions
of $x^i$ alone. But, when (i) either Killing vector field vanishes at at least
one point of your space-time and (ii) the three-form
generated by the outer product of both Killing fields and the Ricci
tensor applied to either of them vanishes, the metric can be put in a
form where the components $g_{\bar{t}\mu}$ and $g_{\bar{\phi}\mu}$ all
vanish. These conditions of Theorem 7.1.1 in \cite{Wald} are met for
the cases of interest, since (i) the rotation vanishes at the axis of
symmetry and (ii) all solutions are supposed to be either vacuum flat or 
vacuum anti-de Sitter, and in either case $R^a_b$ is zero or
proportional to the identity.

On these space-times, one can choose the coordinates $x^i$ in the
surfaces orthogonal to the curves parametrized by $\bar{t}$ and
$\bar{\phi}$, so
that the interval element takes the form:
\be
ds^2=-V\,d\bar{t}^2+2W\,d\bar{t}\,d\bar{\phi}+X\,d\bar{\phi}^2+
g_{ij}dx^i\, dx^j   
\ee
where $V$, $W$, $X$ and $g_{ij}$ are functions of $x^i$. The
function 
\be
\rho^2=VX+W^2
\ee
parametrizes the volume element of the $\bar{t}-\bar{\phi}$ plane and
its vanishing 
is signal that at least one linear combination of the Killing vector
field has vanishing norm, {\it i. e.}, signals the existence of a {\it
  Killing horizon}, which appears in all known black hole solutions
which display an event horizon. We will assume that $\nabla_a\rho\neq
0$, and, for now, the rest of the coordinates $x^i$ constant along the
integral curves of $\nabla^a\rho$, so that $g_{\rho i}=0$. The line
element for a generic space-time allowing for two commuting Killing
vector fields is then 
\be
ds^2=-\frac{\rho^2}{X}d\bar{t}^2+X(d\bar{\phi} -\omega
d\bar{t})^2+g_{\rho\rho} d\rho^2 +g_{ij}dx^i\, dx^j, 
\label{metric}
\ee
where $\omega=-W/X$. 

This form of the metric above reduce to a variety of metrics studied
in the Kerr-CFT correspondence. For instance, the Kerr solution
in Newman coordinates are obtained for
\be
\label{eq:kerr}
\begin{gathered}
\rho^2=\Delta\sin^2\theta, \quad\quad X=\sin^2\theta
\frac{(\tilde{r}^2+a^2)^2- \Delta
  a^2\sin^2\theta}{\tilde{r}^2+a^2\cos^2\theta}, \\
\omega = \frac{2M\tilde{r}a}{\Delta (\tilde{r}^2+a^2\cos^2\theta)}
\quad\quad \Delta = r^2-2Mr+a^2.
\end{gathered}
\ee
Note that the horizon is situated at $\Delta=0$ and that the
transverse coordinate $\theta$ differs from the usual Kerr solution by
a $\rho$-dependent term, so that we have $g_{\rho\theta}=0$, rather
than $g_{\tilde{r}\theta}=0$. 

\section{The Near Horizon limit}

We will now study (\ref{metric}) close to the horizon $\rho^2=0$. In
the general case, the functions $X$, $\omega$ all have finite limits
there. They define respectively the radius of the orbits of
$\partial/\partial \phi$ and the angular velocity on the horizon. The
component $g_{\rho\rho}$ merits some extra attention: it is the norm
of the vector $\partial/\partial \rho$, defined to be the conjugate
vector to the gradient $(d\rho)_a=\nabla_a\rho$. Therefore
$(\nabla^a\rho) (\nabla_a\rho)=(g_{\rho\rho})^{-1}$. Define
$\chi^a=( \partial/\partial t)^a -\omega(\partial/\partial \phi)^a$
the Killing vector field which does vanish at the horizon
$\rho^2=0$. The zeroth law of black hole thermodynamics state that the
function 
\be 
\kappa^2 = -\frac{1}{4}\frac{\nabla_a(\chi^b\chi_b)
  \nabla^a(\chi_c\chi^c)}{\chi^d\chi_d}
\label{defkappa}
\ee 
is constant at the horizon
and there it defines the {\it surface gravity} for static (stationary)
black holes. Note also that $\omega$ is constant at $\rho^2=0$ and
likewise defines the {\it angular velocity} $\Omega_H$ of the
horizon. Using the functions $V$, $W$ and $X$, we can write
$\chi_a\chi^a$ as 
\be
\chi^a\chi_a=-V-2\frac{W}{X}W+\frac{W^2}{X^2}X=-\frac{\rho^2}{X}.  
\ee
The definition of $\kappa$ will result on the following expression for
$g_{\rho\rho}$, 
\be
(g_{\rho\rho})^{-1}=X\kappa^2-\frac{\rho^2}{2X^2}\nabla_a X\nabla^a X
+\frac{\rho}{X^2}\nabla_a\rho\nabla^a X.
\label{omega}
\ee

The expression (\ref{omega}) for $g_{\rho \rho}$ above shows that, if the
horizon has finite surface gravity, the scaling limit of the metric
(\ref{metric}) will show nothing out of the ordinary. If, however,
$\kappa\rightarrow 0$, the black hole becomes extremal and
$g_{\rho\rho}$ diverges at small $\rho$ like ${\cal O}(\rho^{-2})$.
Note that, assuming that $X$ is finite at $\rho=0$, the last term
in (\ref{omega}) also gives a contribution of ${\cal O}(\rho^{2})$ to
$(g_{\rho\rho})^{-1}$. We will then assume that
\be
g_{\rho\rho}=\frac{A}{\rho^{2}}+{\mathscr O}(\rho^0),\quad\quad
\omega=\Omega_H+\bar{\Omega}\rho+{\mathscr O}(\rho^2),
\label{falloff}
\ee
for $A$ and $\bar{\Omega}$ functions of $x^i$.

The dependence of $A$ in the transverse coordinates can
be removed by a redefinition of $\rho$ and $x^i$. Using the
liberty to redefine $\rho$ by a function of $x^i$, one can make
\be
\rho\rightarrow \Phi\rho,\quad\quad 
x^i\rightarrow x^i+ A^i
\ee
such that
\be
g_{ij}\frac{\partial A^i}{\partial \rho}\frac{\partial 
A^j}{\partial \rho}  
=\frac{\rho^2}{X}\left(\Phi^2-A\right) \quad \text{ and }
\quad
g_{ij}\frac{\partial A^i}{\partial \rho}  \frac{\partial A^j}{\partial
  x^k}=-\frac{A}{X \Phi\rho}\frac{\partial\Phi}{\partial
  x^k}. 
\ee
The result is a metric of the form
\be
ds^2=\frac{\Phi^2}{X}\left(-\rho^2 d\bar{t}^2+\frac{d\rho^2}{\rho
    ^2}\right) 
+X\left(d\bar{\phi}-\omega d\bar{t}  
\right)^2 +\bar{g}_{ij}dx^idx^j+\ldots
\ee
where the ellipsis include the subleading terms of
$g_{\rho\rho}$. Taking the scaling limit 
\be
\bar{t}=\lambda^{-1} t,\quad\quad \rho=\lambda  r,\quad 
\quad \bar{\phi}=\phi-\Omega_H\lambda^{-1} t,
\quad\quad \lambda\rightarrow 0
\label{scalinglimit}
\ee
of this metric will yield one of the type ${\rm AdS_2}\times X$
\be
ds^2_{\rm nhe}=\Omega^2\left(-r^2 dt^2+\frac{dr^2}{r^2}
  +\Lambda^2\left(d\phi-r\,dt 
\right)^2 \right)+\tilde{g}_{ij}dx^idx^j
\label{nearhorizon}
\ee
with the subscript ``nhe'' stands for ``near-horizon limit of the
extremal black hole'', and where $\Omega$, $\Lambda$ and
$\tilde{g}_{ij}$ are computed at the horizon and, although may depend on
the other angular variables $x^i$, do not of course depend on $r$
anymore.  Furthermore, $r_0$ is now a constant.

The near horizon metric (\ref{nearhorizon}) displays two new
symmetries. One symmetry is
\be 
r\rightarrow cr,\quad\quad t\rightarrow \frac{t}{c},
\ee
generated by the vector field
\be
\xi_3^a=r \left(\frac{\partial}{\partial
    r}\right)^a-t 
\left(\frac{\partial}{\partial t}\right)^a. 
\ee
Another symmetry is generated by the vector field 
\be
\xi_+^a=(r^{-2}+t^2) \left(\frac{\partial}{\partial
    t}\right)^a-2rt 
\left(\frac{\partial}{\partial r}\right)^a
-\frac{\beta}{r} \left(\frac{\partial}{\partial
    \phi}\right)^a,
\ee
with  $\beta$ a constant. These vector fields will, along with
$\psi^a=(\partial/\partial\phi)^a$ and 
$\xi_-^a=(\partial/\partial t)^a$ form a full ${\rm
  SL(2,\mathbb{R})\times U(1)}$ symmetry.

The role of the scaling transformation (\ref{scalinglimit}) is to decouple the
degrees of freedom at the horizon to the ones external to the Black
Hole region. At first this effect may seem at odds with the fact that
(\ref{scalinglimit}) is a change of coordinates, and hence a ``pure
gauge'' transformation.  Schematically, the transformation can be cast
as
\be
g_{ab}^{\rm nhe}=\lim_{\beta\rightarrow \infty}
\exp\left({-\beta\lie{\xi_3}}\right)\hat{K}g_{ab},
\label{eq:scale-transf}
\ee
where the effect of $\hat{K}$ is to take $\bar{\phi}+\Omega_H\bar{t}$
to $\phi$, the affine parameter to the Killing vector field that
becomes null at the horizon. This transformation changes dramatically
the asymptotic characteristics of the metric, which come from being
asymptotically flat to asymptotically anti-de Sitter. Therefore it
cannot be considered a true diffeomorphism. We will have more to say
along these lines in the next section.

From the definition above it is also obvious that the dilation operator
$\xi_3$ will be a symmetry of $g_{ab}^{\rm nhe}$. That leads to the
question of why the procedure doesn't work with metrics away from
extremality, like, for instance, Schwarzschild. The answer is that,
when the temperature of the horizon is not zero, the limit simply does
not exist. Indeed, as one inspects some components of the metric, like
(\ref{omega}), one sees that, unless $\kappa=0$, the relevant term of
the interval element blows up in the limit $\lambda\rightarrow\infty$.

Decoupling limits like (\ref{scalinglimit}) are at the heart of the
gauge-gravity duality, and the holographic view of the global
symmetries relates scale transformations to the renormalization group
flow. Thus, the appearance of a dilation symmetry at the horizon
points to the fact that we are dealing with an infrared fixed point of
the dual theory. It would be interesting to find explicit examples of
such dual theories and their infrared fixed point counterparts.

\section{The Boundary Conditions}

As we saw in the last section, the structure of the near horizon
metric is universal and is a direct result of the vanishing of the
temperature of a Killing horizon. As the relevant quantities are along
the $t$, $\phi$ and $r$ direction, one can wonder whether the
Brown-Henneaux technique \cite{Brown1986b} will help us count
gravitational states. At the core of the technique is the notion that
some coordinate changes alter the asymptotic charges, like the mass or
the angular momentum, and therefore fail to be true diffeomorphisms of
the solution.

For asymptotically flat solutions of theories whose gravitational
sector reduces at low energy to the Einstein-Hilbert Lagrangian, these
charges are given by
\be
M=-\frac{1}{8\pi}\int_{S}\varepsilon_{a_1\ldots
  a_{n-2}bc}\nabla^b\xi^c
,\text{ and }J=-\frac{1}{16\pi}\int_{S}\varepsilon_{a_1\ldots
  a_{n-2}bc}\nabla^b\psi^c,
\label{charges}
\ee
where $S$ is a ``sphere at infinity''. The expressions above allow for
proofs of the laws of the thermodynamics of black holes. By using
Stokes' theorem on the Komar formula for the asymptotic mass above,
one arrives at
\be
M=2\int_\Sigma \xi^e {R_e}^d\, \epsilon_{a_1\ldots a_{n-1}d}
-\frac{1}{8\pi}\int_{\cal H}\varepsilon_{a_1\ldots
  a_{n-2}bc}\nabla^b\xi^c.
\label{smarr}
\ee
This relates the total mass of the space-time with quantities computed
at the horizon of the black hole ${\cal H}$. Indeed, by writing the
time translation Killing vector field as $\chi^a-\Omega_H\psi^a$,
where $\chi^a$ is the Killing vector associated with the horizon, and
$\psi^a$ the one related to rotations, one can show that this surface
term is 
\be
-\frac{1}{8\pi}\int_{\cal H}\varepsilon_{a_1\ldots
  a_{n-2}bc}\nabla^b\xi^c =
\frac{1}{4\pi}\kappa A + 2\Omega_H J_H
\ee
with $\Omega_H$ the ``angular velocity'' of the horizon and $J_H$ the
angular momentum of the black hole. Given a metric variation
$\delta g_{ab}=\gamma_{ab}$ satisfying the linearized equations of
motion $\delta R_{ab}=0$, the change in the total mass is given by
\cite{Carter1973}: 
\be
\delta M = -\frac{1}{8\pi}\int_{S}\varepsilon_{a_1\ldots
  a_{n-2}bc}\xi^b\nabla_d(\gamma^{cd}-g^{cd}\gamma),
\label{1stlaw}
\ee
which is essentially the variation of the bulk term of (\ref{smarr}),
using the usual formulas for the variation of the Ricci
tensor\footnote{\footnotesize See, for instance, (7.5.14) in
\cite{Wald}.}. In the
case where the linearized equations of motion are satisfied, the
domain of integration can be again brought to the horizon:
\be
\delta M = -\frac{1}{8\pi}\int_{S}\varepsilon_{a_1\ldots
  a_{n-2}bc}\xi^b\nabla_d(\gamma^{cd}-g^{cd}\gamma)
=-\frac{1}{8\pi}\int_{\cal H}\varepsilon_{a_1\ldots
  a_{n-2}bc}\xi^b\nabla_d(\gamma^{cd}-g^{cd}\gamma).
\ee
The evaluation of these quantities at the horizon are tied to the
variation of the intensive quantities $\kappa$ and
$\Omega_H$. Evaluating it \cite{Bardeen1973}, one arrives at the Black
Hole Gibbs relation:
\be
\delta M = -\frac{1}{4\pi}A\delta\kappa-2J_H\delta \Omega_H.
\ee

Given a metric variation resulting from the application of a
diffemorphism $\gamma_{ab}=2\nabla_{(a}\eta_{b)}$, one usually expects
the asymptotic charges to remain the same. In fact, these operations
are the gravitational analogue of gauge transformations and as such
cannot alter the value of gauge-invariant observables. However, some
gauge transformations, dubbed large, do change asymptotic holonomies
and hence can alter boundary conditions like the values of the
charges. Perhaps the most common case of the appearance of those are
in instantons in non-abelian gauge theories. In \cite{Brown1986b} the
authors noticed that these also arise in three-dimensional
gravity with negative cosmological constant, which can be
seen as a Chern-Simons gauge theory. There, the appropriate gauge
group is comprised by local isometries of the metric, ${\rm
  SL(2,\mathbb{R})}\times {\rm SL(2,\mathbb{R})}$. These gauged
transformations are locally diffeomorphisms, but may not satisfy
boundary conditions that keep the asymptotic structure intact. As
large gauge transformations change the value of observables like the
total mass and angular momentum, they should be seen as global
symmetries rather than spurious transformations usually associated
with gauge invariance \cite{Banados:1998gg}. Being global symmetries,
these large gauge transformation do fulfill the bootstrap program to
be used to organize the spectrum and allow for the counting of
states. In the case of Brown and Henneaux \cite{Brown1986b}, the global algebra
can be
written in terms of two copies of Virasoro generators and, after
assuming unitarity and modular invariance, character formulas
 \cite{Cardy1986} can be used -- sucessfully --
to account for the Bekenstein-Hawking formula for the black hole
entropy.

One sees then that the interpretation of the entropy of the black hole
as large gauge transformations is intimately tied to the choice
of boundary conditions. The idea is to allow for boundary conditions to
the gauge fields that allow for changing of the asymptotic charges. On
the other hand, too lax a falloff condition would allow for infinite
changes in those charges. This situation is also undesirable for the
purpose of counting states while keeping thermodynamical quantities
fixed.

In \cite{Guica:2008mu} a proposal
for suitable boundary conditions was made, based on the variation of
general formulas developed by Barnich and Brandt
\cite{Barnich:2001jy}. The boundary conditions were given in terms of the
near-horizon limit coordinates (\ref{nearhorizon}) which cloud their
interpretation as modifiers of the asymptotic charges of the full
space. The conditions are $g_{ab}=\bar{g}_{ab}+\gamma_{ab}$ where
$\bar{g}_{ab}$ is the metric associated with the interval element of
``global'' ${\rm AdS}_2\times X$ 
\be 
ds^2=\Omega^2\left(-(1+\varrho^2)
  d\tau^2+ \frac{d\varrho^2}{1+\varrho^2} +
  \Lambda^2\left(d\varphi-\frac{\varrho}{\varrho_0} d\tau \right)^2
\right)+\tilde{g}_{ij}dx^idx^j
\label{nearhorizon1}
\ee
 and $\gamma_{ab}$ are terms vanishing at the conformal
boundary $\varrho=\infty$ like
\be
h_{\mu\nu}=\left(
\begin{array}{cccc}
{\cal O}(\varrho^2) & {\cal O}(1) & {\cal O}(\varrho^{-2})  \\
h_{\phi t} & {\cal O}(1) & {\cal O}(\varrho^{-1}) \\
h_{r t} & h_{r\phi} & {\cal O}(\varrho^{-3}) \\
\end{array}
\right)
\label{boundaryconditions}
\ee
and terms depending on the angular variable $g_{i\mu}={\cal
  O}(\varrho^{-1})$ except $g_{ir}={\cal O}(\varrho^{-2})$.

These boundary conditions are rather odd from the near-horizon, {\it
  i.e.}, the ${\rm AdS}_2$ point of view. In fact, the conditions do
not even maintain the ``asymptotic triviality'' of the metric, which
can be seen from the fact that $h_{\tau\tau}$ is of the same order in
$\varrho$ as $g_{\tau\tau}$. 

These boundary conditions can however be rather
straightforwardly derived from the requirement that the {\it full
  space} modification of the asymptotic charges to be finite. The
relation between charges in the full Kerr metric (\ref{metric}) and in
the near horizon limit (\ref{nearhorizon}) is surprisingly direct. As
a matter of fact, the relation between them is the scaling limit
(\ref{scalinglimit}), which, for finite $\lambda$, is just a change of
coordinates. Moreover, quantities like the volume element do not
change under scaling, and the only perceptible difference stems from
the transformation of the time coordinate. Therefore,
\be
\label{eq:scaling-prescription}
M=\lambda M^{\rm nhe}_\lambda
\ee
where $M^{\rm nhe}_\lambda$ refers to the near horizon geometry
capped at small, but finite, $\lambda$. It is clear then, that while
the metric obtained at the limit $\lambda\rightarrow 0$ is perfectly
reasonable, the charges computed in this limit may be infinite and
still be related to finite changes of the total mass of the Kerr
space-time.  

The near horizon limit is implemented at the metric level, as an
expansion over $\lambda$:
\be
g_{ab}=\bar{g}_{ab}+\lambda h_{ab}+\ldots
\ee
and substitution $g_{ab}=\bar{g}_{ab}$ valid insofar as the
corrections $\lambda h_{ab}$ are small. This fail to be the case at
large $r$, or $\varrho$. We will then consider the near-horizon
geometry up to scales of order $r\propto \lambda^{-1}$, and define
$M^{\rm  nhe}_\lambda$ as the volume integral up to those
scales, {\it i. e.}, integrated on a asymptotic surface $S(\lambda)$
with $\varrho\simeq \lambda^{-1}$. This ensures that the volume term in
the definition of the
Komar mass:
\be
2\int_\Sigma \xi^e {R_e}^d\, \epsilon_{a_1\ldots a_{n-1}d}
\ee
is finite even in the $\lambda\rightarrow 0$ limit. 

Inspecting the formula for the variation of the total mass
(\ref{1stlaw}) written in terms of the ``${\rm nhe}$'' geometry,
\be
\delta M =
-\frac{\lambda^2}{8\pi}\int_{S(\lambda)}\varepsilon_{a_1\ldots
  a_{n-2}bc}\xi^b\nabla_d(h^{cd}-g^{cd}h),
\label{1stlawnhe}
\ee
one sees that the leading order terms in $\varrho$ stems
from the multiplicative terms involving the variation of the metric
and the Christoffel symbols of (\ref{nearhorizon1}): Following the
condition that the variation should be finite in the
$\lambda\rightarrow 0$ limit, one arrives at the boundary conditions
for the near horizon metric as (\ref{boundaryconditions}). Therefore,
even changing dramatically the boundary conditions of the near horizon
metric, these metric variations still yield sensible asymptotic
charges in the full Kerr geometry. One should also remark that, from
the discussion in the last Section, the fact that the near-horizon
geometry is still ${\rm AdS}_2\times X$ after the metric variation means
that one is keeping extremality. Thus one does not need to enforme
similar constraints on the variation of the angular momenta. 

To summarize, the boundary conditions (\ref{boundaryconditions}) stem
from the physical requirement that changes on the near extremal metric
still allow for finite changes in the asymptotic charges of the full
metric. As stated, the purpose of those boundary conditions is to
provide the requirements on the space-times which will be counted by the
asymptotic symmetries
\be
\xi_\epsilon=\epsilon(\phi)\partial_\phi-\varrho\epsilon'(\phi)
\partial_\varrho .
\ee
The surprising feature of such solution is that, despite its
simplicity, it manages to be a orbit on the space of global gauge
transformations, {\it i. e.}, each variation counts exactly one
diffeomorphically inequivalent metric. Without a microscopic theory,
one cannot prove it so at this point. We hope to return to this point
in the future.

\section{Central Charges}

In this section, we will describe the procedure to obtain
asymptotic charges
algebra with respect to the boundary conditions obtained in previous
section.
This algebra will be a conformal algebra with a central charge on which
we will focus. As stated in the preceeding sections, the central charge
is the key ingredient on character formulas for the number of states at
a given level. The entropy obtained in this manner is, surprinsingly, 
equal to Bekenstein-Hawking entropy for an extremal Kerr black hole.

Most of this Section will follow the guidelines of \cite{Guica:2008mu},
and, since the general form of the metric (\ref{nearhorizon}) poses no
additional challenge, will be somewhat schematical. By the prescription
(\ref{eq:scaling-prescription}) for the relation between Kerr mass and a
divergent ``nhe''
mass, we have obtained the boundary conditions
(\ref{boundaryconditions}). These conditions allow one to
obtain the the generators of the asymptotic symmetries
\begin{eqnarray}
\label{eq:as-sym-witt}
\xi_\epsilon&=&\epsilon(\phi)\partial_\phi-\varrho\epsilon'(\phi)
\partial_\varrho \\
\zeta_t&=& \partial_t. \label{eq:as-sym-t}
\end{eqnarray}
Observe that $\zeta_t$ generates time translations. Furthermore, since
we are treating $\phi$ as a periodic, we may set
$\epsilon_n(\phi)=-e^{-in\phi}$, for $n$ an
integer. Therefore, we have an infinite number of generators for the
asymptotic
symmetry generated by $\xi_\epsilon$, i.e., $\xi_n=\xi(\epsilon_n)$. These
infinite generators are closed into a ${\rm Diff~S}^1$ algebra:
\begin{equation}
	i\left[ \xi_n,\xi_m\right]=(m-n)\xi_{m+n}.
\end{equation}
For $n=0$, $\xi_0$ generates a simple rotational isometry, i.e., $U(1)$.

The next step is to obtain asymptotic charges associated with asymptotic
symmetries, and then a Dirac algebra for these asymptotic charges. Conserved
asymptotic charges associated with asymptotic symmetries $\zeta$ are
defined by \cite{Barnich:2001jy}:
\begin{equation}
	Q_\zeta[\bar{g}]=\frac{1}{8\pi}~\int_{S}
k_\zeta[h;\bar{g}],
\end{equation}
where $S=\partial \Sigma$ is the boundary of a codimension one
hypersurface, and
$h_{ab}$ encodes boundary conditions for the field (background)
$\bar{g}_{ab}$. Furthemore, $k_{\zeta}$ is an asymptotically
conserved
$n-2$-form given by $k_\zeta[h;\bar{g}]=k^{[ab]}_\zeta[h;\bar{g}]
(d^{(n-2)}x)_{ab}$, such that
\begin{equation}
	k^{[ab]}_\zeta[h;\bar{g}]=\frac{\sqrt{-\bar{g}}}{16\pi}\left(
\zeta_c \bar{\nabla}_\sigma
H^{cdab}+\frac{1}{2}H^{cdab}\partial_{c}\zeta_d \right),
\end{equation}
where the covariant derivative $\bar{\nabla}$ is with respect to background
$\bar{g}_{ab}$, and $H^{cdab}$ is a background tensor
with
indices symmetries similar to a Riemman tensor\footnote{See eqs. (6.19) and
(6.20) in \cite{Barnich:2001jy}, which we reproduce here.}
\begin{eqnarray}
	H^{cdab}[h;\bar{g}]&=&-\hat{h}^{cb}\bar{g}^{da}-\hat{h}^{da}
\bar{g}^{cb}+\hat{h}^{ca}\bar{g}^{db}+\hat{h}^{db}
\bar{g}^{ac} \\
\hat{h}_{ab}&=& h_{ab}-\frac{1}{2}\bar{g}_{ab} h.
\end{eqnarray}
Obviously, in the above expression indices are raised and lowered with the
background metric $\bar{g}_{ab}$.

The Dirac bracket algebra for these asymptotic charge may display a central
charge. Formulae for this algebra and the central charges are
\begin{equation}
	\left\{ Q_{\zeta_1},Q_{\zeta_2} \right\}=\delta_{\zeta_1}
Q_{\zeta_2}+K_{\zeta_1,\zeta_2}
\end{equation}
and
\begin{equation}
	  \label{eq:central-charge1}
	K_{\zeta_1,\zeta_2}=\frac{1}{8\pi} \int_{S}
k_{\zeta_1}[\mathcal{L}_{\zeta_2} \bar{g},\bar{g}],
\end{equation}
where $\mathcal{L}_{\zeta_2}$ is a Lie derivative along $\zeta_2$. We
draw attention to the fact \cite{Barnich:2001jy} that
$K_{\zeta_1,\zeta_2}=-K_{\zeta_2,\zeta_1}$.
We now use the above formalism for the asymptotic symmetries given
(\ref{eq:as-sym-witt}) and (\ref{eq:as-sym-t}). For the time translation
$\zeta_t$, as discussed in \cite{Guica:2008mu}, its associated asymptotic charge
has to vanish identically. Therefore, this implies further supplementary
boundary conditions. Now, the boundary conditions (\ref{boundaryconditions})
does satisfy $Q_{\zeta_t}\equiv 0$ \footnote{see footnote 10 in
\cite{Guica:2008mu}.}.

For asymptotic charges associated to generators of the algebra
(\ref{eq:as-sym-witt}), one obtain a Virasoro algebra with central charge
computed from the integral (\ref{eq:central-charge1}). If we define Virasoro
generators $L_n$ as $L_n=Q_{\xi_n}+\frac{3J}{2}\delta_{n,0}$. The by Dirac's
quantization procedure ($\{\cdot,\cdot \}\to-i[\cdot,\cdot]$),
then
\begin{equation}
	[L_m,L_n]=(m-n)L_{m+n}+\frac{c_L}{12} m(m^2-1)\delta_{m+n,0},
\end{equation}
with
\begin{equation}
         \label{eq:central-charge-J}
	c_L=12 J.
\end{equation}

\subsection{Frolov Temperature}

The Frolov-Thorne temperature for Kerr black hole is obtained
analogously as the
Hawking temperature for Schwarzschild black hole. There, one first define a
Hartle-Hawking vacuum outside the black hole to be a mixed state
$\rho=e^{-E/T_H}$.
For the extremal Kerr back hole, this definition needs to be worked on
since $T_H=0$. The solution for the mixed state stems from the
existence of another thermodynamic potential related to
the angular momentum $J$ of the black hole. Thus, from the first Law,
\begin{equation}
	T_H \delta S=\delta M-\Omega_H \delta J.
\end{equation}
In the extremal case, $T_H \delta S=0$, then $\delta M=\Omega_H \delta J$.
Following  \cite{Hartman:2008pb}, one has in this case that the entropy
variation is given by $\delta S=\delta J/T_L$. Now, the entropy for the
extremal Kerr black Hole is $S=2\pi J$. Therefore we set the
Frolov-Thorne temperature to be $T_L=1/2\pi$. One notes at this point
that this is the inverse of the periodicity of the coordinate $\phi$
defined at (\ref{scalinglimit}). In this fashion, it arises naturally
as the periodicity required so that the metric has no conical
singularity after analytic continuation to Euclidian time.

In the spirit of Section 4, where we have obtained boundary conditions
(\ref{boundaryconditions}), we may obtain the same value for $T_L$ by a
semi-classical argument. The first step is to define an analogue of
Hartle-Hawking vacuum for the Kerr Black Hole, {\it i.e.}, Frolov-Thorne
vacuum. For
that, one consider, for simplicity, a scalar quantum field $\Phi$ in a
classical four dimensional Kerr background. Now, expand this scalar
field in eigenstates of asymptotic energy $\omega$ and angular momentum
$l,m$ as
\begin{equation}
	\Phi(x)=\sum_{\omega,l,m} \Phi_{\omega,l,m} e^{-i\omega \bar{t} +
	im\bar{\phi}}~F_{l}(\rho,x^i).
\end{equation}
One now has to trace out the interior of the black hole to obtain the vacuum as
a density matrix in the above basis as $e^{-\frac{\omega-\Omega_H m}{T_H}}$.
However, since $T_H=0$, this expression for density matrix
is not suitable. One way of solving of this problem is to rewrite the
coordinates in
the scaled form (\ref{scalinglimit}), and take the limit 
$\lambda\to 0$ after the procedure. In the coordinates
(\ref{scalinglimit}), the field is written as
\begin{equation}
	\Phi(x)=\sum_{\omega,l,m} \Phi_{\omega,l,m} e^{-i\omega\frac{t}{\lambda}
	+ im (\phi-\Omega_H \frac{t}{\lambda})}~F_{l}(\lambda r,x^i),
\end{equation}
which can be suitably recast as
\begin{equation}
	\Phi(x)=\sum_{\omega,l,m} \Phi_{\omega,l,m} e^{-i(E_R t + J
\phi)}~F_{l}(\lambda r,x^i),
\end{equation}
with
\begin{equation}
	J\equiv m, ~\quad\quad ~E_R\equiv
\frac{\omega-\Omega_H}{\lambda}.
\end{equation}
Once again, one trace out the interior of the black hole. In this case, we obtain
\begin{equation}
       \label{eq:dens-mat}
	\rho = e^{-\frac{E_R}{T_R}-\frac{J}{T_L}},
\end{equation}
with
\begin{equation}
	T_L= \frac{\Omega_H}{2\pi\omega} \quad
\text{and}\quad T_R=\kappa,
\end{equation}
in units of $M$.

Now, we take the limit $\lambda\to 0$. If $E_R$ is not zero, then
the extremality limit $\kappa\rightarrow 0$ would make the density
matrix (\ref{eq:dens-mat}) vanish. Then there would only be pure
states. If one, on the other hand, enforces the extremal limit in terms
of the double limit $\lambda \to 0$ and $\omega-\Omega_H\to0$, then
$E_R$ still tends to zero, but the density matrix is now non-vanishing,
since
\begin{equation}
	T_L=\frac{1}{2\pi}.
\end{equation}

\subsection{Entropy}

The Bekenstein-Hawking entropy for Kerr black hole is obtained plugging
the central charge (\ref{eq:central-charge-J}) into Cardy-Verlinde
formula
\begin{equation}
	S_{\textrm{micro}}=\frac{\pi^2}{3}~c_L~T_L.
\end{equation}
If we use the Frolov-Thorne temperatute for extremal Kerr black
hole
(\ref{eq:kerr}), then
\begin{equation}
	S_{\textrm{micro}}=2\pi J\equiv S_{\textrm{Kerr}}.
\end{equation}
Therefore, via the counting of microstates encoded in central charge, $c_L$, we
obtained a thermodynamic entropy $S_{\textrm{Kerr}}$.

We should however observe that use of Cardy-Verlinde formula here has to be done
\emph{cum grana salis}. Indeed, Cardy-Verlinde formula is valid,
provided the theory is unitary and modular invariant, when $T\gg c$.
But here $T_L\ll c_l$, or at best of the same order. The same happens in
many other situations, {\it e. g.}, in Strominger-Vafa work. There
are some attempts to understand this numerical coincidence in the
literature, still lacking a good explanation. This example of
decoupling limit between the near horizon degrees of freedom is unique
in the fact that it does not seem to rely on the usual theorems of
nonrenormalizability of supersymmetric theories. It would be then 
interesting to explicit the exact property that protects the sector of
the dual theory in the renormalization group flow.


\section{Conclusions}

In this letter we studied the underlying space-time mechanism behind
the Kerr/CFT correspondence. We took an space-time approach, trying to
understand the emergence of the conformal symmetry and to elucidate the
mechanism behind the coincidence of Cardy's formula and the
Bekenstein-Hawking entropy. We revisited, for the first part, that the
conformal symmetry is a direct consequence of the existence of a
zero-temperature Killing horizon. Heuristically, any such horizon will
have $\partial/\partial\rho$ have its norm diverging as
(\ref{falloff}), and the ${\rm SL(2,\mathbb{R})}$ structure will appear.

For the second part, we have given some arguments that the
Bekenstein-Hawking entropy is again counting inequivalent metrics. The
Kerr/CFT correspondence helps us count metrics which change global
charges despite being generated by diffeomorphisms. When one restricts
such metrics to the near horizon limit, one arrives at the boundary
conditions proposed by \cite{Guica:2008mu}. Again heuristically, the
boundary conditions are such that allows for finite displacements of
the global charges of the entire black hole, even though the near
horizon strutucture changes considerably. We then review some features
of these global diffeomorphisms, which displaying a
central charge in their representation in a extremal black hole
background. 

We conclude posing the question of what feature of the horizon is
necessary for the statement that the sector of the dual theory is
decoupling. This idea has been around some time under the name of
``attractor mechanism'' \cite{Sen:2007qy} (see also
\cite{Astefanesei:2006dd}, and \cite{Astefanesei:2009sh} for a
discussion of the purported dual CFT). Albeit full supersymmetry
seems unnecessary, the algebraic structure of the black hole like the
existence of a twice repeated null principal vector should play a role.
One should remember at this point that the existence of special
algebraic structures is sufficient to prove that the spacetime is an
exact target space for string theory, or, more generically, that the
spacetime is free from quantum corrections to the geometry. The
existence of a covariantly constant null vector field, for instance,
falls into both categories \cite{Tseytlin:1992pq,Coley:2008th}. We hope
to say more along those lines in the near future.

\section*{Acknowledgements}

We would like to thank Mirjam Cvetič, Brian Wecht and
especially Luciano Barosi for comments and suggestions. We would also
like to thank Francisco Brito, Álvaro Ferraz for support. BCdC
 thanks the ICCMP-UnB, where part of this work was conducted. ARQ
acknowledges partial support of FINATEC and also CNPq under grant no.
307760/2009-0.


\end{document}